\documentclass[twoside]{article}

\usepackage{lipsum} 

\usepackage[sc]{mathpazo} 
\usepackage[T1]{fontenc} 
\usepackage{microtype} 

\usepackage[hmarginratio=1:1,top=32mm,columnsep=20pt]{geometry} 
\usepackage{multicol} 
\usepackage[hang, small,labelfont=bf,up,textfont=it,up]{caption} 
\usepackage{booktabs} 
\usepackage{float} 
\usepackage{hyperref} 

\usepackage{lettrine} 
\usepackage{paralist} 
\usepackage{graphicx}
\graphicspath{ {/} }
\usepackage{abstract} 

\usepackage{titlesec} 
\renewcommand\thesection{\Roman{section}} 
\renewcommand\thesubsection{\Roman{subsection}} 
\titleformat{\section}[block]{\large\scshape\centering}{\thesection.}{1em}{} 
\titleformat{\subsection}[block]{\large}{\thesubsection.}{1em}{} 

\usepackage{fancyhdr} 
\title{\vspace{-15mm}\fontsize{24pt}{10pt}\selectfont\textbf{Android Tapjacking Vulnerability}} 

\author{
\large
\textsc{Benjamin Lim (A0100223)}\\[2mm] 
\normalsize National University of Singapore \\ 
\normalsize limbenjamin@u.nus.edu \\ 
}
\date{April 22, 2015}

\begin{document}

\maketitle 

\thispagestyle{fancy} 


\begin{abstract}

\noindent Android is an open source mobile operating system that is developed mainly by Google. It is used on a significant portion of mobile devices worldwide. In this paper, I will be looking at an attack commonly known as tapjacking. I will be taking the attack apart and walking through each individual step required to implement the attack. I will then explore the various payload options available to an attacker. Lastly, I will touch on the feasibility of the attack as well as mitigation strategies.

\end{abstract}


\begin{multicols}{2} 

\section{Introduction}

\lettrine[nindent=0em,lines=2]{T}he tapjacking attack basically tricks the user into tapping on an object in the background layer by clever positioning of a foreground layer that is not tappable. Hence, any user touches will be applied onto the background layer which is not visible to the user. It is essentially a delivery mechanism and the payload can be customised by the attacker. The exploit is payload and aspect ratio specific, therefore the exploit code will need to be modified depending on the payload desired by the attacker as well as the target device's aspect ratio. The attack is also limited by the screen real estate of the device, I will be elaborating more on that in the section on developing the application.


\section{Exploiting the vulnerability}

\subsection{Payload Selection}

The first step in developing the exploit will be to choose a payload. For this walkthrough, I will be using the application installer payload. We will need to note down the location and number of taps a user would make in order to install an application. In the case of Google Play, the steps are as follows.

\begin{enumerate}
\item Open the App detail \footnote{We can access the app detail page directly through market:// url. Hence, we do not need to search for the app.} page of the target app
\item Tap Install
~\\
~\\
\includegraphics[width=180px,keepaspectratio]{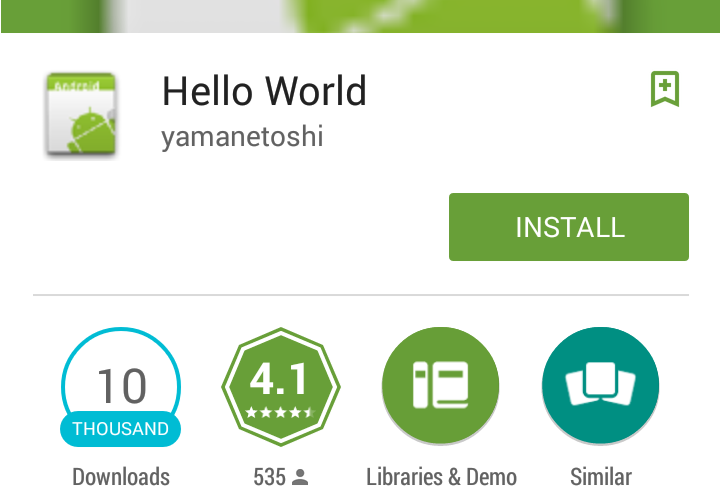}
\\
\item Tap Accept
~\\
~\\
\includegraphics[width=180px,keepaspectratio]{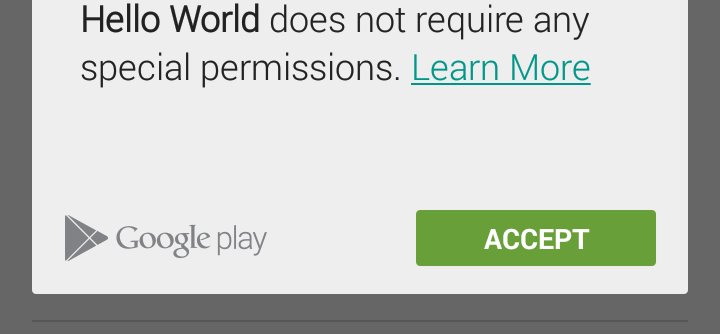}
\\
\end{enumerate}

\pagebreak
\end{multicols}
\includegraphics[width=\textwidth,keepaspectratio]{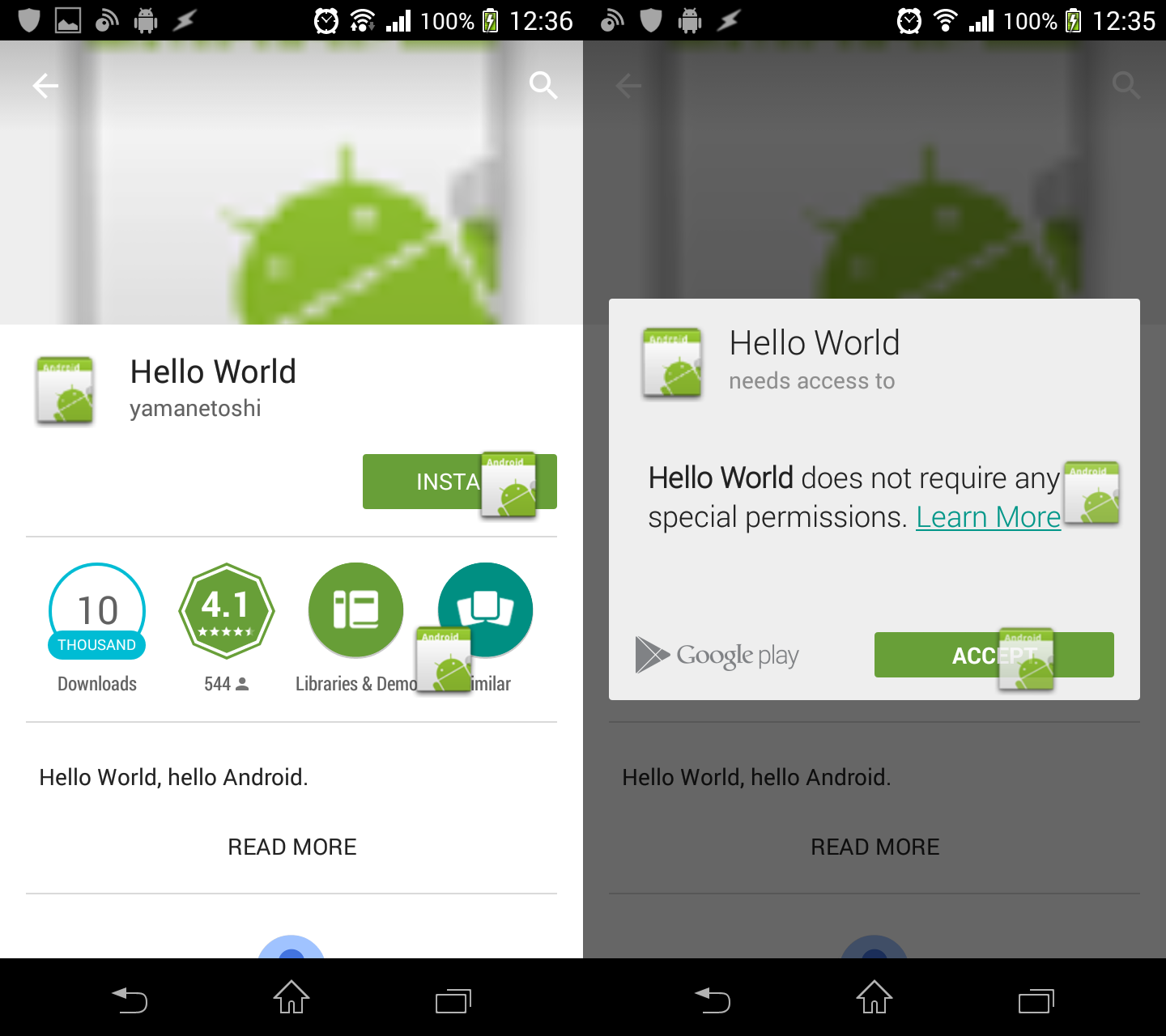}
\\
\begin{multicols}{2}

\subsection{Developing the application}

Once the desired payload and steps has been identified, we can move on to developing the application. We would need to create a toast activity and have the image overlay the buttons which need to be pressed. Toasts are normally used to display short text notifications and any taps will be filtered down to the background layer. Positioning of the toast has to be done by trial and error. We will want to use density independent pixels (dp) when specifying the position so that the exploit code will work on devices with different resolutions but same aspect ratios. 
\\
\\The images have to be placed such that no image overlaps a tappable area of any previous screen. E.g. The image for the install button has been shifted to the left slightly so it does not overlap the "Learn More" link in the permissions page. This minimises the probability of the exploit failing. Thus the attack in practical is limited to 2 to 3 clicks at most due to limited screen real estate. Furthermore, the attack will also be unlikely to work if the size of the button is too small as it will be difficult as the victim might not be able to tap the exact spot. 

\pagebreak
The next step would involve setting the toast to repeat on a loop so that is always displayed on the screen and set the background of the toast to white so as to obscure the target application.
~\\
~\\
\includegraphics[width=180px,keepaspectratio]{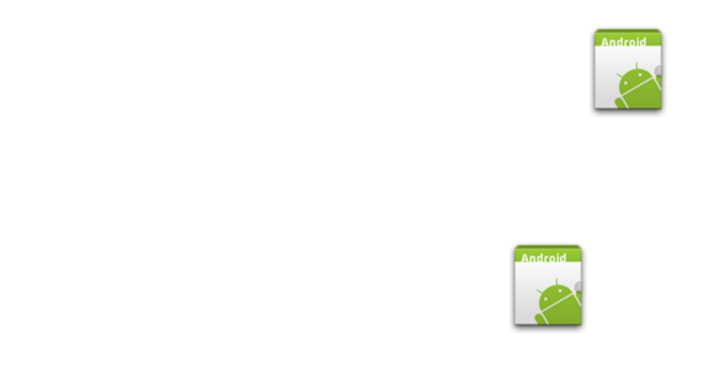}
~\\
~\\
At this point, we might want to include baits promising the user an incentive if they tap on the image repeatedly. We are now done with the development of the exploit and it can be packaged and installed on the target device via an appropriate method.


\section{Attack Impact}

As mentioned in the introduction section, the tapjacking attack is a delivery mechanism, hence its impact would depend on the payload.  

\subsection{Installer Payload}
Assuming that the attacker chose to use the installer payload, he would be able to perform a privilege escalation through the stealthy installation of a second app which requires multiple permissions that the user did not agree to. The exploit app itself does not require any permissions.
~\\
~\\
\includegraphics[width=180px,keepaspectratio]{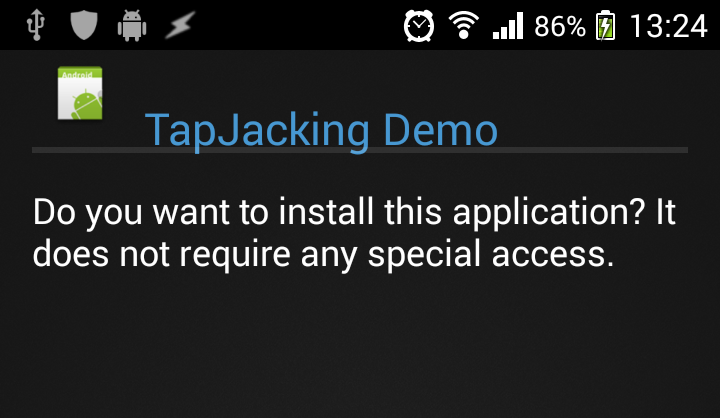}
~\\
~\\
The second app will most likely request the following core permissions.
\begin{enumerate}
\item RECEIVE\_BOOT\_COMPLETED - Allows the attacker to start a service in the background whenever the phone is restarted. Thus the user does not even need to run the application.
\item INTERNET - Allows upload of data on the phone to the attacker's server
\item ACCESS\_NETWORK\_STATE - The attacker might want to upload data only when WiFi is active so as not to use up too much quota and raise suspicions.
\end{enumerate}

Depending on the attacker's motive, he can make use of any of the following permissions (ACCESS\_FINE\_LOCATION, CAMERA, RECORD\_AUDIO, READ\_CALENDAR, READ\_CALL\_LOG, READ\_CONTACTS, READ\_SMS, READ\_EXTERNAL\_STORAGE... ) to compromise the privacy of the target user. 

There are a few tactics an attacker can use to conceal the attack from the user. One of the methods involves removing the app icon from the launcher and can be achieved by replacing "android.intent.category.LAUNCHER" in the manifest file with "android.intent.category.DEFAULT". Therefore, the user will not be able to locate the app when he swipes through the list of installed apps on the launcher. The second method is to use a generic name such as "Android Update Service" or "Bluetooth Connection Helper". On encountering such an application, a user will likely assume that the application is part of the Android operating system and will ignore the application.

\pagebreak

\subsection{Alternative Payload}

\subsubsection{URL based Payload}

Apart from the installer payload, which is triggered by opening a "market://" URL in a webview, other URL based payloads include a "http(s)://" and a "tel://" payload. The HTTP payload will allow an attacker to open any URL inside a webview. The webpage could contain a full screen button which could trigger a file download or run code which exploits a vulnerability in the webview container. However, the tapjacking application will need the permission to access the internet which could raise user's suspicions.  
\\
\\The "tel://" payload will cause the user to silently dial a number in the background. It does not have such a high impact on security and the worst that could happen would be that the attacker programmed the payload with a premium number and the user would be left with a higher phone bill than expected. 

\subsubsection{Other Intent based Payload}

Apart from URLs, an attacker can also use intents to launch other activities. For example, the Settings activity can be called up using the following snippet of code "new Intent( android.provider.Settings.ACTION\_SETTINGS)". With the Settings activity in the background, the attacker can then trick the user into performing various actions ranging from switching on and off Wifi and Bluetooth to allowing installation of apps from unknown sources. 
\\
\\The attacker can also use intents to launch third party applications using the following code snippet "getPackageManager().getLaunchIntentForPackage( "com.bank.app"); The impact of the attack would depend on the app in question. Needless to say, the attacker would need to ensure that the target user has the target application installed and must be familiar with the various activities and layout of the target application. This variation of the attack is thus one of the most difficult to pull off. 


\section{Attack Feasibility}

\begin{enumerate}
\item Exploitability - Proof of Concept
\item Impact - High
\item Complexity - Very High
\item Overall - Low
\end{enumerate}

Only proof of concept code is available at the moment. Thus an attacker will need to know basic android development in order to write or modify the code needed to exploit the vulnerability. As of now, there does not exist any tool which would automate the development of such an app when fed a payload.
\\
\\The impact of the attack is (potentially) high and depends on the type of payload. In the case of the installer payload, an attacker would be able to access the call information, SMSes, location, files on SD card, camera and microphone, completely compromising the privacy of the user. Hence, the impact is relatively severe.
\\
\\Complexity is very high because the attacker has first got to convince the user to install the application. He then has to convince the user to comply with the instructions and tap repeatedly on the images. Lastly, there is a substantial chance of failure especially if the user's taps are not accurate. 
\\
\\In summary, the attack is not feasible because it requires the attacker to be skilled enough to write custom code and the user to be gullible enough to follow through with instructions. The attack is also not scalable as it only works on devices of a specific aspect ratio. A skilled attacker would be able to compromise phones in masses using easier techniques. Therefore, this attack is not feasible and likely only used in a targeted attack.

\pagebreak

\section{Mitigation Strategies}

According to an unofficial source \cite{1}, the tapjacking vulnerability was claimed to have been patched in Android version 4.0.3. However, I have successfully carried out this attack on my phone which is running Android 4.3. I am unable to ascertain if this is because the manufacturer of my phone has not applied the patch in their images or whether the patch does not exist.
\\
\\Developers can set the filterTouchesWhenObscured property to true or override the onFilterTouchEventForSecurity method. Setting the property to true is the declarative security method and will ignore all taps when the app is not in the foreground. Overriding the method is the programmatic security approach and gives the developer more flexibility. He can choose to ignore or to process the taps based on certain conditions. i.e. if the app was in the foreground within the last 5 seconds. Given that even Google Play itself is vulnerable, it is unlikely that many developers practice either one of the methods above.
\\
\\This is little that users can do to guard themselves against a tapjacking attack. But in general, users should try not to download obscure apps or download apps from third party app stores. They should look out for suspicious behaviour such as unsolicited app installs and practice common sense.


\section{Conclusion}

I have walked through the process of planning and developing an application that exploits the tapjacking vulnerability. Even though there is not much an android user can do to protect himself from such an attack, there is little cause for concern as the attack is not feasible to pull off. Nevertheless, android users should still adopt good security practices to thwart other attacks out there. Lastly, developers should also play a more active role in ensuring that their applications are safe from such attacks.  


\section{Acknowledgements}

Part of the code was shamelessly taken from nVisium's tapjacking proof of concept \cite{2}. The code was then revised for a more updated version of the android SDK and customised for the aspect ratio of my phone. I then stripped out some of the features so I could demonstrate how tapjacking works in the background.

A copy of my application code can be found at https://github.com/limbenjamin/tapjacking and is open sourced under the MIT license. 

\end{multicols}

\pagebreak


\end{document}